\documentclass{jnmauth}

\begin{document}
\JNM{161}{174}{16}{28}{03}

\runningheads{A.\ Aste and R. Vahldieck}
{Time-Domain Simulation of the Full Hydrodynamic Model}

\title{Time-Domain Simulation of the Full Hydrodynamic Model}

\author{Andreas Aste\affil{1}, R\"udiger Vahldieck\affil{2}}

\address{\affilnum{1}\ Institute for Theoretical Physics,
Klingelbergstrasse 82, 4054 Basel, Switzerland\\
\affilnum{2}\
Laboratory for Electromagnetic Fields and Microwave Electronics,\\
Gloriastrasse 35, 8092 Z\"urich, Switzerland}

\corraddr{Institute for Theoretical Physics,
Klingelbergstrasse 82, 4054 Basel, Switzerland}

\footnotetext{Work supported by the Swiss National
Science Foundation, Project no. 2100-57176.99}

\received{29 April 2002}
\revised{28 September 2002}
\accepted{19 December 2002}

\begin{abstract}
A simple upwind discretization of the highly
coupled non-linear differential equations which define the
hydrodynamic model for
semiconductors is given in full detail. The hydrodynamic model
is able to describe inertia effects which
play an increasing role in different fields of opto- and microelectronics.
A silicon $n^+-n-n^+$-structure is simulated, using
the energy-balance model and the full hydrodynamic model. Results 
for stationary cases are
then compared, and it is pointed out
where the energy-balance model, which is implemented in
most of today's commercial
semiconductor device simulators, fails to describe accurately the
electron dynamics. Additionally, a GaAs $n^+-n-n^+$-structure is simulated
in time-domain in order to illustrate the importance of inertia effects
at high frequencies in modern submicron devices.
\end{abstract}


\section{Introduction}
Our paper emerges from the the fact that today's
submicron semiconductor devices
are operated under high frequencies and strong
electric fields.
Information transmission using electromagnetic waves at
very high frequencies will have
a direct impact on how we design active and passive components
in different fields of micro- and optoelectronics.
In such cases, quasi-static semiconductor device models
like the energy-balance model (EBM) are no longer adequate.
Especially in GaAs and related materials used for high-speed device design,
inertia effects play an important role since the impulse and energy
relaxation times of the electron gas are close to the picosecond range.

The most elaborate and practicable
approach for the description of charge transport
in semiconductors used for device simulation would be the Monte Carlo (MC)
method \cite{0}. The advantage of this technique is a complete picture of
carrier dynamics with reference to microscopic material parameters,
e.g. effective masses and scattering parameters. But the method must be
still considered as very time consuming and hence not economical
to be used by device designers.

Besides the simplest concept which is the traditional
drift-diffusion model (DDM),
there is a much more rigorous approach to the problem, namely the so-called
hydrodynamic model (HDM). This model makes use of electron temperature (or
energy density) as additional quantities for the description of charge
transport. Starting from the Boltzmann equation, Blotekjaer \cite{1}
and many others presented a derivation of such equations for the
first time. The physical parameters used in the HDM
can be obtained from theoretical considerations or MC simulations.
A simplified version of the HDM is the so-called
energy-balance model (EBM).

In the first part of this work, we give a short definition
of the charge transport models. We will illustrate how the different
models and their parameters can be related to each other. In a second part,
we give a simple discretization scheme for the full hydrodynamic model. In the
last part, we compare the different models for the case of a
submicron silicon ballistic diode (an $n^+-n-n^+$ structure) and a
Gallium Arsenide ballistic diode.

\section{The HDM for silicon}

Since we will
illustrate the HDM for the case of an n-doped ballistic diode where
the contribution of electron holes to the current transport is
negligible, we will only discuss the charge transport models for electrons.
Generalization of the models to the case where both charge carriers
are present is straightforward.

The four hydrodynamic equations for parabolic energy bands are

\begin{equation}
\frac{\partial n}{\partial t}+\vec{\nabla} \vec{j}=0\quad , \label{cont}
\end{equation}

\begin{equation}
\frac{\partial \vec{j}}{\partial t} + (\vec{\nabla}\vec{j})\vec{v}+
(\vec{j}\vec{\nabla})\vec{v}=-\frac{e}{m}n\vec{E}-
\frac{e}{m} \vec{\nabla} \Bigl( \frac{nkT}{e} \Bigr)-\frac{\vec{j}}{\tau_p}
\quad , \label{ib}
\end{equation}

\begin{displaymath}
\frac{\partial \omega}{\partial t}+\vec{\nabla}(\vec{v} \omega) =
\end{displaymath}
\begin{equation}
-en\vec{v}\vec{E} - \vec{\nabla}(nkT\vec{v})-\vec{\nabla}(-\kappa
\vec{\nabla}T) - \frac{\omega-\frac{3}{2}nkT_L}{\tau_\omega}
\quad ,\label{eb}
\end{equation}
and
\begin{equation}
\vec{\nabla} (\epsilon \vec{\nabla}\Phi)=e(n-N_D) \quad , \label{poi}
\end{equation}
where $n$ is the electron density, $\vec{v}$ the drift velocity of the
electron gas, $e>0$ the elemental charge, $\vec{E}=
-\vec{\nabla} \Phi$ the electric field,
$\Phi$ the quasi-static electric potential,
T the electron gas temperature, $\omega$ the electron
energy density, $\kappa$ the thermal conductivity, $\epsilon$ is the
dielectric constant, and $N_D$ is the density of donors.

The particle current density $\vec{j}=n\vec{v}$ is related to the current
density $\vec{J}$ by the simple formula $\vec{J}=-e\vec{j}$.

Eq. (\ref{cont}) is simply the continuity equation which expresses
particle number conservation. Eq. (\ref{ib}) is the so-called
impulse balance equation, eq. (\ref{eb}) the energy balance equation
and eq. (\ref{poi}) the well-known Poisson equation.

We will solve the hydrodynamic equations for $n,\vec{j},\omega$ and
$\Phi$. To close the set of equations, we relate the electron energy
density to the thermal and kinetic energy of the electrons
by assuming parabolic energy bands
\begin{equation}
\omega=\frac{3}{2}nkT+\frac{1}{2}nmv^2 \quad.
\end{equation}
In fact, we already assumed implicitly parabolic bands for the
impulse balance equation, which is usually given in the form
\begin{equation}
\frac{\partial \vec{p}}{\partial t} + (\vec{\nabla}\vec{p})\vec{v}
+(\vec{p}\vec{\nabla})\vec{v}=-en\vec{E} -\vec{\nabla}
(nkT)-\frac{\vec{p}}{\tau_p} \quad ,
\end{equation}
i.e. we replaced the electron impulse density by the particle current
density by assuming $\vec{p}=mn\vec{v}$, where $m$ is a {\em{constant}}
effective electron mass.

The impulse relaxation time $\tau_p$ describes the impulse loss of
the electron gas due to the interaction with the crystal, the energy
relaxation time $\tau_\omega$ the energy transfer between the electron gas
with temperature $T$ and the crystal lattice with temperature $T_L$.
$\tau_p$ and $\tau_\omega$ are usually modelled as a function of
the total doping density $N_D+N_A$ where $N_A$ is the density of
acceptors, the lattice temperature $T_L$,
the electron temperature $T$ or alternatively
the mean energy per electron $\omega/n$.

A simplification of the full HDM is
the energy-balance model.
In the EBM, the convective terms of the impulse balance equation are skipped.
The energy balance equation is simplified by the assumption that
the time derivative of the mean electron energy
$\partial \omega/\partial t$ is small compared to the other terms and that
the kinetic part in $\omega$ can also be neglected, i.e.
\begin{equation}
\omega=\frac{3}{2}nk T \quad.
\end{equation}
This non-degenerate approximation which avoids a description
by Fermi integrals
is justified for the low electron
densities in the relevant region of the simulation examples, where
velocity overshoot can be observed.

The energy balance equation then becomes
\begin{displaymath}
\vec{\nabla}(\frac{5}{2}nkT\vec{v})-\vec{\nabla}(\kappa \vec{\nabla}T) =
\end{displaymath}
\begin{equation}
-en\vec{v}\vec{E} - \frac{\frac{3}{2}nk(T-T_L)}{\tau_\omega} \quad ,
\end{equation}
and the impulse balance equation becomes the current equation
\begin{equation}
\vec{j}=-\frac{e}{m}\tau_p n \vec{E}-\frac{e}{m} \tau_p \vec{\nabla}
\Bigl( \frac{nkT}{e} \Bigr) \quad .
\end{equation}
Continuity equation and Poisson equation are of course still valid in
the EBM.
Neglecting the time derivative of the current density is equivalent
to the assumption that the electron momentum is able to adjust itself
to a change in the electric field within a very short time. While this
assumption is justified for relatively long-gated field effect
transistors, it needs to be investigated for short-gate cases.

A further simplification of the EBM leads to the drift-diffusion model.
The energy balance equation is completely removed from the set of
equations, therefore it is no longer possible to include the
electron temperature $T$ in the current equation. $T$ is simply
replaced by the lattice temperature $T_L$. 
Therefore the DDM consists of the continuity equation, the Poisson
equation and the current equation
\begin{equation}
\vec{j}=-\frac{e \tau_p}{m} n \vec{E} -\frac{e \tau_p}{m} \Bigl( \frac{kT_L}
{e} \Bigr) \vec{\nabla} n \quad , \label{ce}
\end{equation}
and it is assumed that the electron mobility is a function of the
electric field.
But at least, the electron
temperature is taken into account in an implicit way: If one considers
the stationary and homogeneous case in the HDM, where spatial and temporal
derivatives can be neglected, one has
for the current equation
\begin{equation}
\vec{j}=n\vec{v}=-\frac{e \tau_p}{m} n \vec{E}
\end{equation}
or
\begin{equation}
\vec{v}=-\frac{e \tau_p}{m} \vec{E} \quad , \label{hom1}
\end{equation}
and the energy balance equation becomes simply
\begin{equation}
-e \vec{v} \vec{E}=\frac{\frac{3}{2}k(T-T_L)+\frac{1}{2}mv^2}{\tau_\omega}
\quad . \label{hom2}
\end{equation}
Combining eq. (\ref{hom1}) and (\ref{hom2}) leads to the relation
\begin{equation}
\frac{e^2}{m}(\tau_p \tau_\omega-\tau_p^2/2) E^2 = \frac{3}{2}k(T-T_L) \quad .
\end{equation}
In our simulations for silicon, we will use the
Baccarani-Wordemann model, which defines the relaxation times by
\begin{equation}
\tau_p=\frac{m}{e} \mu_0 \frac{T_0}{T} \label{relax1}
\end{equation}
\begin{eqnarray}
\tau_w=\frac{m}{2e}\mu_0 \frac{T_0}{T}+\frac{3}{2} \frac{k}{e v_s^2} \mu_0 
\frac{TT_0}{T+T_0}=
\nonumber \\
\Bigl( \frac{1}{2}+\frac{3k}{2mv_s^2}\frac{T^2}{T+T_0} \Bigr)
\tau_p \quad . \label{relax2}
\end{eqnarray}
$v_s$ is the saturation velocity, i.e. the drift velocity of the electron gas
at high electric fields. $\mu_0$ is the low field mobility, which depends
mainly on the lattice temperature and the total doping density.

For the sake of completeness, we mention that
inserting the expressions for the relaxation times
into eqs. (\ref{hom1}) and (\ref{hom2}) leads to the
$E(T)$-relation
\begin{equation}
E^2=\frac{v_s^2}{\mu_0^2} \Biggl[ \Bigl(\frac{T}{T_L}\Bigr)^2-1 \Biggr]
\end{equation}
and the electron mobility $\mu=(e/m)\tau_p$ is given by
\begin{equation}
\mu(E)=\frac{e \tau_p(E)}{m} =
\frac{\mu_0}{\sqrt{1+\Bigl(\frac{\mu_0 E}{v_s}\Bigr)^2}} \quad . \label{ct}
\end{equation}
This is the well-known Caughey-Thomas mobility model \cite{6}.
It has the important property that $v(E) \sim \mu_0 E$ for
$E << \frac{vs}{\mu_0}$ and $v \sim  v_s$ for $E>> \frac{v_s}{\mu_0}$.

The EBM has the big advantage that it includes the electron temperature $T$,
such that the electron temperature gradient can be included in the
current equation, and the mobility can be modelled more accurately
as a function of $T$.

An expression is needed for the thermal conductivity of
the electron gas, which stems from theoretical considerations
\begin{eqnarray}
\kappa=(5/2+r)n \frac{k^2 \mu(T)}{e} T \quad .  \label{heat}
\end{eqnarray}
Several different choices for r can be found in the literature,
and many authors
\cite{2,3} even neglect heat conduction in their models.
But Baccarani and Wordeman point out in \cite{4} that neglecting
this term can lead to nonphysical results and mathematical instability.
Although their work is directed to Si, their remarks should be equally
valid for GaAs since the equations have the same form in both cases.
We will present a GaAs MESFET simulation comparable to the
one of Ghazaly {\em{et al.}} \cite{3} in a forthcoming paper,
but with heat conduction included.
The best value for $r$ appears to be $-2.1$ for silicon at 300 K, according
to comparisons of hydrodynamic and MC simulations of the ballistic diode
\cite{8}.

\section{Discretization scheme}

Today, many elaborate discretization methods are available for
the DDM equations or EBM equations. The well-known Schar\-fet\-ter-Gum\-mel
method \cite{9} for the DDM makes use of the fact that the current
density is a slowly varying quantity. The current equation is then
solved exactly under the assumption of a constant current density over
a discretization cell,
which leads to an improved expression for the current density than it is
given by simple central differences. It is therefore possible to
implement physical arguments into the discretization method.
Similar techniques have been worked out for the EBM \cite{10}.
But due to the complexity of the HDM equations, no
satisfactory discretization methods which include physical input
are available for this case. For the one dimensional case, this
is not a very big disadvantage, since the accuracy of the
calculations can be improved by choosing a finer grid, without
rising very strongly the computation time.

Therefore, we developed a shock-capturing
upwind discretization method, which has the
advantage of being simple and reliable.
For our purposes, it was sufficient to
use a homogeneous mesh and a constant time step.
But the method can be generalized
without any problems to the non-homogeneous case. Also a generalization
to two dimensions causes no problems, and we are currently studying
two dimensional simulations of GaAs MESFETs.

The fact that the discretization scheme is fully explicit should not
mislead to the presumption that it is of a trivial kind. In fact,
stabilizing a fully explicit discretization scheme for such a highly
non-linear system of differential equations like the HDM is a difficult task,
and a slight change in the discretization strategy may cause instabilities.
Therefore, naive application of the upwind method does not lead to
the desired result.
The order how the different quantities are updated is also of crucial
importance for the maximal timesteps that are allowed. The timesteps can be
enhanced by using an implicit scheme, but only at the cost of an
increased amount of computations needed for the iterative numerical solution
of the implicit nonlinear equations.

The device of length $l$ is decomposed into $N$ cells $C_i$ of equal length
$\Delta x=l/N$. 'Scalar' quantities like the electron density $n_i$,
$(i=1,...N)$,
the potential $\Phi_i$ and the electron energy density $\omega_i$ are
thought to be
located at the center of the cells, whereas 'vectorial' quantities
like the particle current $j_{i+1/2}$, $(i=1,...N-1)$ and the electric
field $E_{i+1/2}=(\Phi_i-\Phi_{i+1})/\Delta x$
are located at the boundaries of the cells. If necessary,
we can define e.g. $E_i$ by
\begin{equation}
E_i=\frac{1}{2}(E_{i-1/2}+E_{i+1/2}) \quad ,
\end{equation}
but a different definition will apply to e.g. $j_i$, as we shall see.

The fundamental variables that we will have to compute at each timestep
are $n_i$, $\Phi_i$, $\omega_i$ (or $T_i$) for $i=2,...N-1$, and $j_{i+1/2}$
for $i=1,...N-1$, if $n_1$, $n_N$, $\Phi_1$, $\Phi_N$, $\omega_1$, and
$\omega_N$ are fixed by boundary conditions. All other variables
used in the sequel should be considered as derived quantities.

The constant timestep $\Delta t$ used in our
simulations was typically of the order of a
few tenths of a femtosecond, and quantities at time $T=t \Delta t$ carry
an upper integer index $t$.

Having calculated $n_i^t$ for a timestep $t$, we define the electron
density at the midpoint by
\begin{equation}
n_{i+1/2}^t=\left\{ \begin{array}{r@{\quad:\quad}l}
\frac{3}{2} n_{i}^t-\frac{1}{2}n_{i-1}^t & j_{i+1/2}^t >0 \\
\frac{3}{2} n_{i+1}^t-\frac{1}{2} n_{i+2}^t & j_{i+1/2}^t < 0
\end{array} \right. \quad ,
\end{equation}
i.e the electron density is extrapolated from neighbouring points
in the direction of the electron flow,
and further
\begin{equation}
v_{i+1/2}^t=j_{i+1/2}^t/n_{i+1/2}^t \quad .
\end{equation}
The upwind extrapolation of the electron density which is
given by the weighting factors $3/2$ and $-1/2$ is improving
the accuracy of the scheme
compared to the usual upwind choice $n_{i \pm
1/2}=n_i$, where simply the neighbouring value in upwind direction
is used.
Analogously we define
\begin{equation}
j_{i}^t=\left\{ \begin{array}{r@{\quad:\quad}l}
\frac{3}{2} j_{i-1/2}^t-\frac{1}{2}j_{i-3/2}^t) & j_{i+1/2}^t >0 \\
\frac{3}{2} j_{i+1/2}^t-\frac{1}{2} j_{i+3/2}) & j_{i+1/2}^t < 0
\end{array} \right. \quad ,
\end{equation}
and
\begin{equation}
v_i^t=j_i^t/n_i^t \quad .
\end{equation}

The discretization of the Poisson equation can be done by central differences.
The continuity equation is discretized as follows:
\begin{equation}
\frac{n_i^{t+1}-n_i^t}{\Delta t} = - \frac{j_{i+1/2}^t-j_{i-1/2}^t}{\Delta x}
\quad , \quad i=2,...N-1 \quad , \label{cons}
\end{equation}
and thus $n_i^{t+1}$ can be calculated from quantities at $T=t \Delta t$.
Eq. {\ref{cons}} defines a conservative discretization, since the
total number of electrons
can only be changed at the boundaries, where electrons may enter or leave
the device. Electrons inside the device
which leave cell $C_i$ at its right boundary enter cell $C_{i+1}$
from the left. The values of $j_{N+1/2}$ and $j_{1/2}$ will not be
needed in our simulations, since we will use boundary conditions
for the electron density which fix $n_1$ and $n_N$.

As a next step we have to discretize the impulse balance equation.
Most of the terms can be discretized by central differences:
\begin{eqnarray}
\frac{j_{i+1/2}^{t+1}-j_{i+1/2}^{t}}{\Delta t}=-\frac{e}{m} n_{i+1/2}^t
E_{i+1/2}^t-
\nonumber \\
\frac{k}{m}(n_{i+1}^t T_{i+1}^t-
n_{i}^t T_{i}^t)/\Delta x-\frac{j_{i+1/2}^t}
{\tau_{p,i+1/2}^t}-(conv)_{i+1/2}^t
\quad , \\
E_{i+1/2}^t=(\Phi_i^t-\Phi_{i+1}^t)/\Delta x \quad ,
\end{eqnarray}
but the convective terms require an upwind discretization
\begin{eqnarray}
(conv)_{i+1/2}^t & = &j_{i+1/2}^t(v_{i+1/2}^t-v_{i-1/2}^t)/\Delta x+
\nonumber \\
& & v_{i+1/2}^t(j_{i+1/2}^t-j_{i-1/2}^t)/\Delta x
\end{eqnarray}
if $j_{i+1/2}^t$ or $v_{i+1/2}^t$ have positive direction
and otherwise
\begin{eqnarray}
(conv)_{i+1/2}^t & =  & j_{i+1/2}^t(v_{i+3/2}^t-v_{i+1/2}^t)/\Delta x+
\nonumber\\
& & v_{i+1/2}^t(j_{i+3/2}^t-j_{i+1/2}^t)/\Delta x \, .
\end{eqnarray}

We observed that the stability of the scheme is improved
for silicon if
the current density is first updated by
\begin{eqnarray}
\frac{\hat{j}_{i+1/2}^{t+1}-j_{i+1/2}^{t}}{\Delta t}=-\frac{e}{m} n_{i+1/2}^t
E_{i+1/2}^t-
\nonumber \\
\frac{k}{m}(n_{i+1}^t T_{i+1}^t-
n_{i}^t T_{i}^t)/\Delta x-\frac{j_{i+1/2}^t}{\tau_{p,i+1/2}^t} \quad ,
\end{eqnarray}
and then $\hat{j}_{i+1/2}^t$ 
is updated by the convective terms, but with $j_{i+1/2}^t$
and $v_{i+1/2}^t$ replaced by the values resulting from
$\hat{j}_{i+1/2}^t$.

The electron temperature is related to the energy density
by the relation $\omega_i^t=\frac{3}{2}nk T_i^t+\frac{1}{2}mn_i^t v_i^{t2}$
and can therefore be regarded as a dependent variable.
The energy balance equation is discretized by defining first
\begin{equation}
\omega_{i+1/2}^t=\left\{ \begin{array}{r@{\quad:\quad}l}
\frac{3}{2} \omega_{i}^t-\frac{1}{2}\omega_{i-1}^t & j_{i+1/2}^t >0 \\
\frac{3}{2} \omega_{i+1}^t-\frac{1}{2} \omega_{i+2}^t & j_{i+1/2}^t < 0
\end{array} \right. \quad ,
\end{equation}
such that $T_{i+1/2}^t$ is also defined by our upwind procedure.
Then the discretization is given by
\begin{eqnarray}
\frac{\omega_i^{t+1}-\omega_i^t}{\Delta t} & = & -en_i^tv_i^tE_i^t-
\frac{\omega_i^t-\frac{3}{2}n_i^tkT_L}{\tau_{\omega,i}^t}
\nonumber \\
& & -\frac{1}{\Delta x}(j_{e,i+1/2}^t-j_{e,i-1/2}^t) \nonumber \\
& & -\frac{1}{\Delta x}(j_{p,i+1/2}^t-j_{p,i-1/2}^t) \nonumber \\
& & -\frac{1}{\Delta x}(j_{h,i+1/2}^t-j_{h,i-1/2}^t) \quad ,
\end{eqnarray}
where we have defined three energy currents
\begin{equation}
j_{e,i+1/2}^t=v_{i+1/2}^t \omega_{i+1/2}^t \quad ,
\end{equation}
\begin{equation}
j_{p,i+1/2}^t=kj_{i+1/2}^t T_{i+1/2}^t \quad ,
\end{equation}
and
\begin{equation}
j_{h,i+1/2}^t=-\kappa (T_{i+1}^t-T_{i}^t)/\Delta x \quad .
\end{equation}

\section{Stationary simulation results}

We simulated an $n^+-n-n^+$ ballistic diode,
which models the
electron flow in the channel of a MOSFET, and exhibits hot electron effects at
scales on the order of a micrometer. Our diode begins with an
$0.1 \, \mu$m $n^+$
"source" region with doping density $N_D=10^{18} \mbox{cm}^{-3}$,
is followed by an $0.1 \, \mu$m
n "channel" region ($N_D=2 \cdot 10^{15} \mbox{cm}^{-3}$),
and ends with an $0.1 \, \mu$m $n^+$ "drain" region
(again $N_D=10^{18} \mbox{cm}^{-3}$).
The doping density was slightly smeared out at the junctions.
We used the following physical parameters for si\-li\-con at $T_L=300 \mbox{K}$
\cite{5}:
The effective electron mass $m=0.26m_e$, where $m_e$ is the electron mass,
$\epsilon=11.7$, and $v_s=1.03 \cdot 10^5 \mbox{m/s}$.
The low field mobility is given by the empirical formula
\begin{equation}
\mu_0(N_D)=\mu_{min}+\frac{\Delta \mu}{1+(N_D/N_{ref})^{0.72}} \quad ,
\end{equation}
\begin{equation}
\mu_{min}=80 \mbox{cm}^2/\mbox{Vs} \quad , \quad \Delta \mu=1430
 \mbox{cm}^2/\mbox{Vs} - \mu_{min} \quad ,
\end{equation}
\begin{equation}
N_{ref}=1.12 \cdot 10^{17} \mbox{cm}^{-3} \quad .
\end{equation}
The temperature dependent mobilities and relaxation times follow
from the low field values according to eqs. (\ref{relax1},\ref{relax2}).

For boundary conditions we have taken charge neutral contacts
in thermal equilibrium with the ambient temperature
at $x=0$ and $x=l=0.3 \, \mu$m, with a bias $V$ across the device:
\begin{equation}
n_1=N_D(0) \quad , \quad n_N=N_D(l) \quad ,
\end{equation}
\begin{equation}
T_1=T_N=T_L \quad , \quad \Phi_1=0 \quad , \quad \Phi_N=V \quad .
\end{equation}
Iinitial values were taken from a simple DDM equilibrium state
simulation.

Stationary results were obtained by applying to the device in thermal
equilibrium a
bias which increased at a rate of typically 1 Volt per picosecond from
zero Volts to the desired final value.
After 6 picoseconds, the stationary state was {\em{de facto}} reached,
i.e. the current density was then constant up to $10^{-4}$\%.

In most cases we used $N=200$ discretization cells,
which proved to be accurate enough, and time steps
$\Delta t$ of the order of a femtosecond. A comparison with simulations with
$N>500$ shows that all relevant quantities do not
differ more than about 5\% from the exact solution.

The computation of a stationary state on a typical modern workstation requires
only few seconds of CPU time, if FORTRAN 95 is used.

\begin{figure}
\centering\includegraphics[width=10 cm]{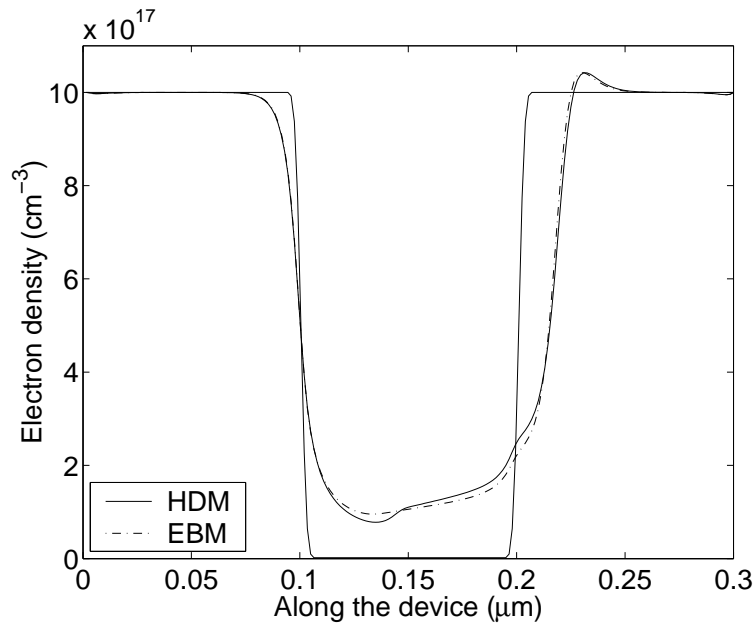}
\caption{Electron density for $V$=1 V (stationary state).
The HDM exhibits more structure than the EBM.
The symmetric curve is the nearly abrupt doping profile.}
\end{figure}

Fig. 1 shows the electron density for the EB and HD
charge transport models for a bias of 1 V.
The choice of 1 V is meaningful since at higher bias ($>2$ V),
the HDM would no longer be applicable or the device
would even be destroyed.

In this paper, solid lines refer
always to the HDM anddashdotted lines to the EBM.
The HDM exhibits more structure than the
EBM. It is interesting to observe that the electron flow becomes
supersonic at $x=0.109 \, \mu$m in the HDM (whereas it remains
subsonic in the DDM). Fig. 2 shows also the soundspeed in the
electron gas calculated from the electron temperature in the HDM,
which is given by $c=\sqrt{kT/m}$ if heat conduction is
included and by $c=\sqrt{5kT/3m}$ otherwise (dashed curve).
In fact, a shock wave
develops in the region where the Mach number $v/c$ is greater than one.
In the DDM, the electron velocity exceeds the saturation
at most by 30\%.
The maximum electron velocity in the
HDM is $2.61 \,  v_s$, in the EBM only $1.87 \, v_s$.

\begin{figure}
\centering\includegraphics[width=10 cm]{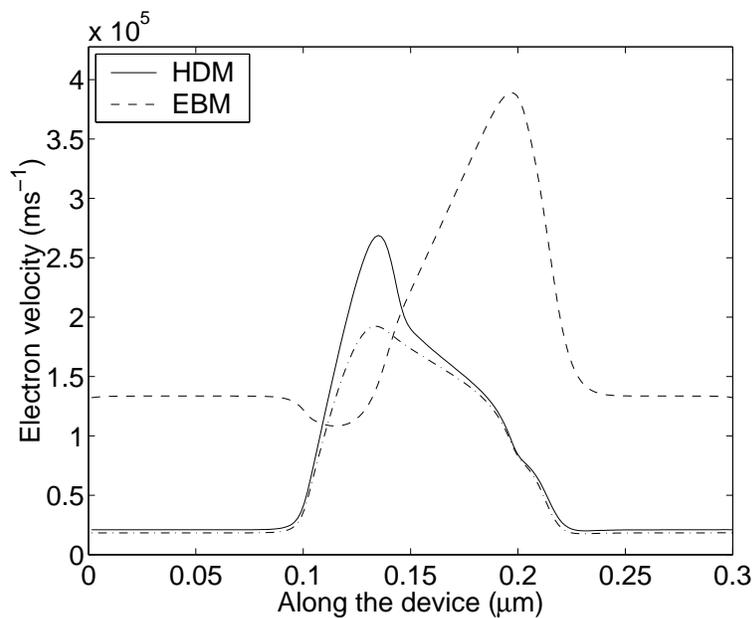}
\caption{Electron velocity for the different charge transport
models ($V$=1 V). Between $x=0.109 \, \mu\mbox{m}$ and $x=0.146 \,
\mu\mbox{m}$ the
HDM electron velocity is supersonic. The dashed curve is the soundspeed.}
\end{figure}
Finally we observe in Fig. 3 that the EBM is able to describe the electron
temperature in an acceptable way. It also predicts the cooling of the
electron gas near $x=0.1 \, \mu \mbox{m}$,
which is caused by the little energy
barrier visible in Fig. 4, where the electric field has a positive value.
The dotted line in Fig. 4 show the electric field in thermal equilibrium.

\begin{figure}
\centering\includegraphics[width=10 cm]{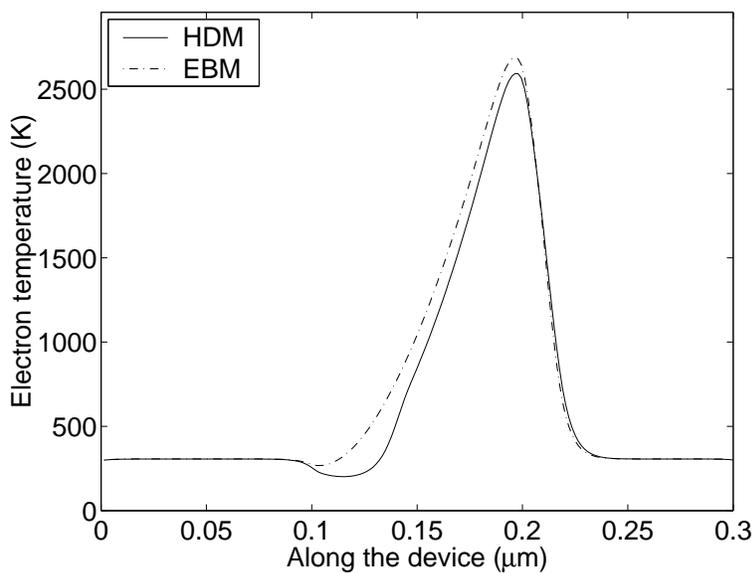}
\caption{Electron temperature for $V$=1 V. HDM and EBM are in good agreement.
(The DDM electron temperature calculated from the electric
field becomes meaningless in this case.)}
\end{figure}

\begin{figure}
\centering\includegraphics[width=10 cm]{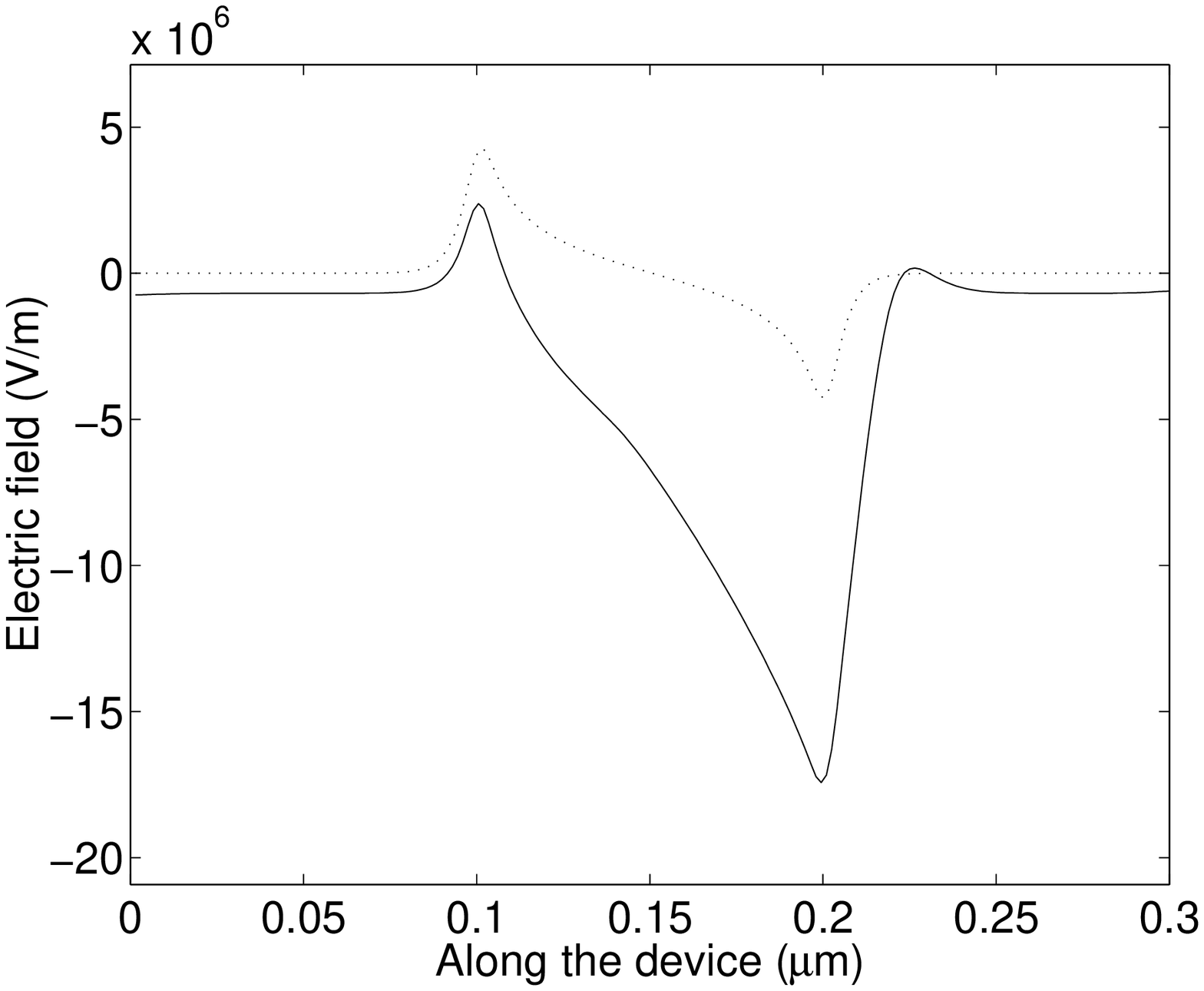}
\caption{Electric field inside the device.}
\end{figure}
But still it is interesting to note that the drift part of the mean
electron energy $\omega/n$ becomes large in a small region around the
drain-source junction, where the electron kinetic energy can be as large
as 58\% of the total energy (Fig. 5).

\begin{figure}
\centering\includegraphics[width=10 cm]{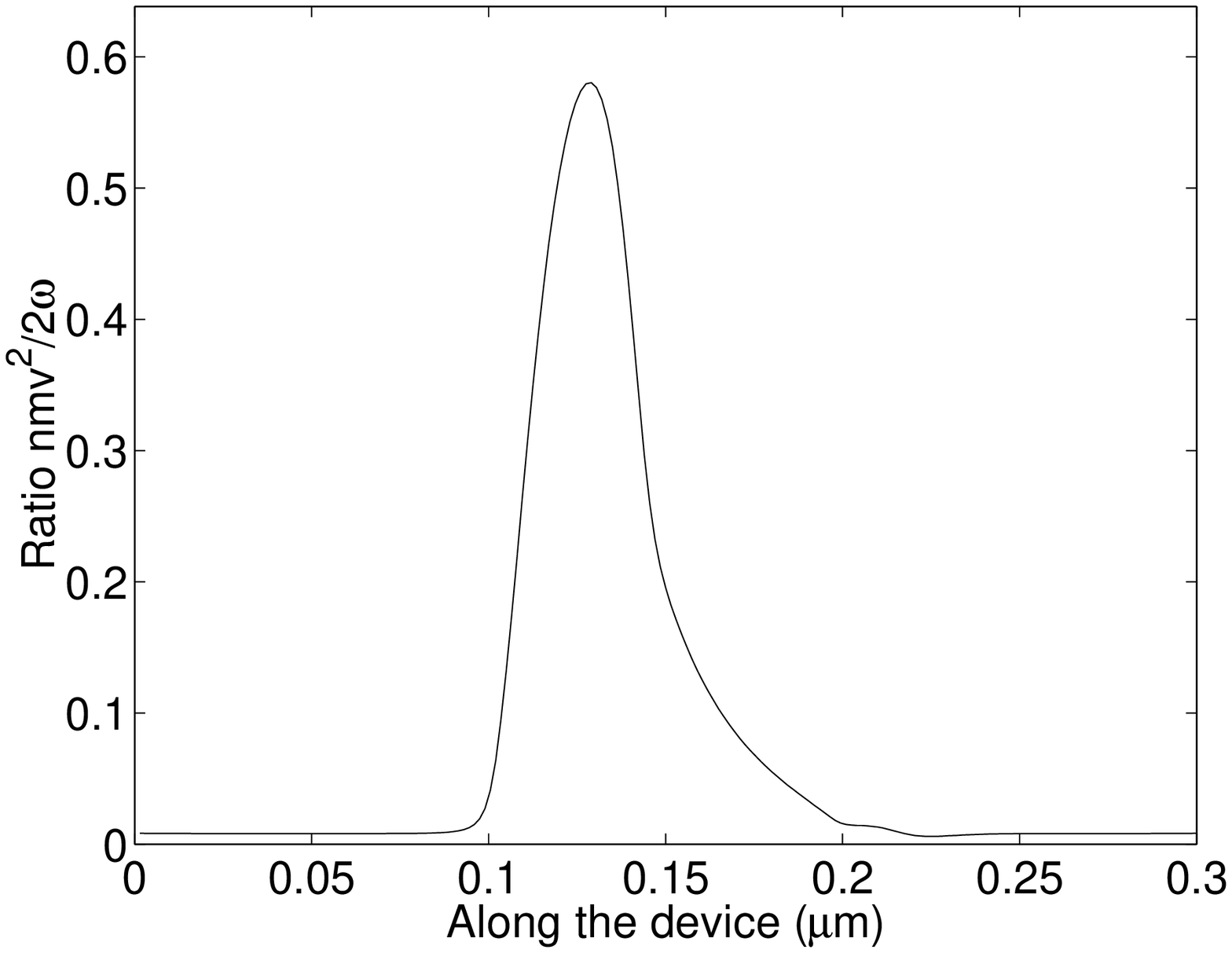}
\caption{Ratio of kinetic drift energy and total energy of electrons for
$V$=1 V.}
\end{figure}

From the point of view of device modeling, the J-V-characteristics
resulting from the three models is of importance (Fig. 6). At low bias, the
electron mobility is in all three models is governed by the
low field mobility $\mu_0$.

\begin{figure}
\centering\includegraphics[width=10cm]{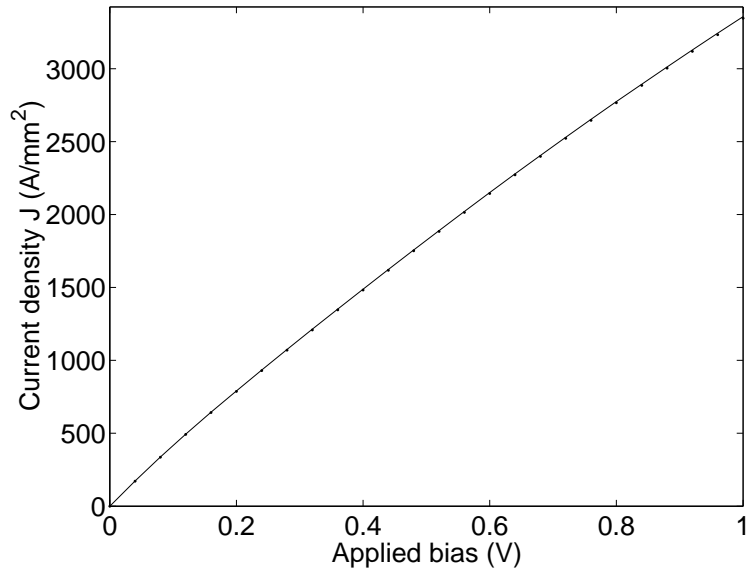}
\caption{$J-V$-characteristics of the device (HDM). The dots represent
data obtained from the EBM.}
\end{figure}
But it is quite
astonishing how the EBM predicts a J-V-curve which is in very good
agreement with the HDM prediction also in the range of higher biases, such
that the two curves in Fig. 6 are nearly undistinguishable.
As we have already mentioned, the EBM does not take inertia effects
into account, which play no role for the stationary case. The
predictive power of the two models is quite different, as we will
see in the next section. But in the stationary case, the difference
in the J-V-characteristics is, roughly speaking, averaged out.

\section{Inertia effects}
For GaAs, the relaxation times are quite high. Therefore, inertia
effects will become important if the applied electric field
changes at a high frequency.
We simulated therefore a GaAs ballistic diode at 300 K.
The diode begins with an $0.2 \, \mu$m $n^+$
source region with doping density $N_D=2 \cdot 10^{17} \mbox{cm}^{-3}$,
is followed by an $0.4 \, \mu$m
n channel region ($N_D=2 \cdot 10^{15} \mbox{cm}^{-3}$),
and ends with an $0.2 \, \mu$m $n^+$ drain region
($N_D=2 \cdot 10^{17} \mbox{cm}^{-3}$). The relevant data like energy-dependent
relaxation times and electron mass were obtained by two-valley MC
simulations, where also the non-parabolicity of the two lowest conduction
band valleys in GaAs was taken into account. For the sake of brevity
we will not go into details here, which will be given in a forthcoming paper
concerning the full hydrodynamic simulation of a GaAs MESFET structure.

In order to show the different behavior of the HDM and EBM in time-domain,
we applied to the 0.1 V pre-biased ballistic diode an additional 0.1 V
pulse of 1 picosecond duration (see Fig. 7).
Fig. 8 shows the particle current density in the exact middle of the device
for both models as a function of time. Whereas the current in the EBM
reacts immediately to the applied field, the current in the HDM shows
relaxation effects. The impulse relaxation time in the channel of the
diode is of the order of 0.3 picoseconds, the energy relaxation time
lies between 0.3 and 1 picosecond.
We emphasize the fact that considering the total current
\begin{equation}
\vec{j}_{tot}=-en\vec{v}+\epsilon_0 \epsilon_r \frac{\partial \vec{E}}
{\partial t}
\end{equation}
does not help; the effect remains. Therefore we must conclude
that the EBM, which is often termed "hydrodynamic model"
in commercial semiconductor device simulators, may lead accidentally
to reasonable (static) characteristics of a device, although the physical
processes inside the device are modelled incorrectly.

\begin{figure}
\centering\includegraphics[width=10 cm]{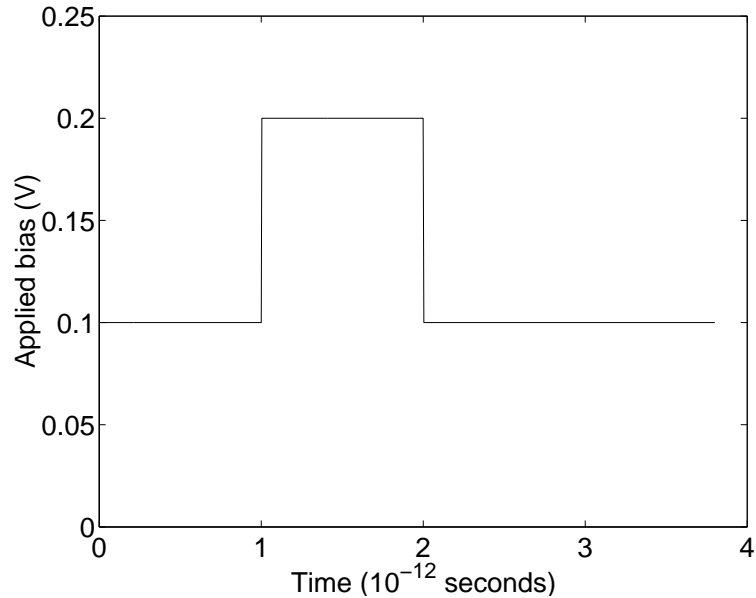}
\caption{Applied bias as a function of time.}
\end{figure}

\begin{figure}
\centering\includegraphics[width=10 cm]{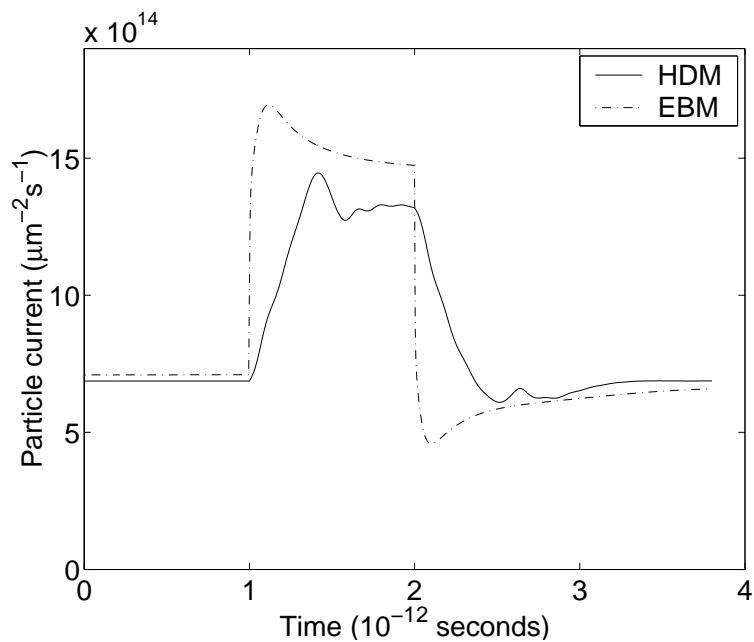}
\caption{Particle current density inside the ballistic diode as a
function of time for the pulse depicted in Fig. 7.}
\end{figure}

\section{Conclusion}
A simple discretization scheme for the hydrodynamic model is
given in full detail which gives
a valuable tool to the practitioner
entering the field of hydrodynamic device modeling.
Comparisons of the different transport models show that the
energy-balance model is capable of describing the behavior of
submicron devices fairly well, but full hydrodynamic simulations are
needed in order to give a satisfactory description of the device
from the physical point of view.
At high frequencies, inertia effects become important in GaAs and
related materials. It is therefore clear that the EBM will be no longer
adequate for simulation of high-speed submicron devices in the
near future.
%

%

\subsection*{Biographies}
{\bf Andreas Aste} received the diploma degree in theoretical
physics from the University of Basel, Basel, Switzerland, in 1993,
and the Ph.D. degree
from the University of Z\"urich, Z\"urich, Switzerland, in 1997. From 1997 to
1998 he was a post doctoral assistant at the Institute for Theoretical
Physics in Z\"urich. From 1998 to 2001
he was a research assistant and Project
Leader in the Laboratory for Electromagnetic Fields and Microwave
Electronics of the Swiss Federal Institute of Technology ETH.
Since 2001 he is working as a researcher at the Institute for
Theoretical Physics at the University of Basel.

\noindent Dr. Aste is a member of the American Physical Society APS.

\vskip 0.4 cm
\noindent {\bf R\"udiger Vahldieck} received the Dipl.-Ing. and Dr.-Ing.
degrees in
electrical engineering from the University of Bremen, Germany,
in 1980 and 1983, respectively.
From 1984 to 1986, he was a Research Associate at the University of Ottawa,
Ottawa, Canada. In 1986, he joined the Department of Electrical and
Computer Engineering, University of Victoria,
British Columbia, Canada, where he became a
Full Professor in 1991. During Fall and Spring 1992-1993, he was visiting
scientist at the Ferdinand-Braun-Institute f\"ur Hochfrequenztechnik in
Berlin, Germany. Since 1997, he is Professor of field theory at the
Swiss Federal Institute of Technology, Z\"urich, Switzerland.
His research interests include numerical methods to model electromagnetic
fields in the general area of electromagnetic compatibility (EMC) and
in particular for computer-aided design of microwave, millimeter wave
and opto-electronic integrated circuits.

Prof. Vahldieck, together
with three co-authors, received the 1983 Outstanding Publication Award
presented by the Institution of Electronic and Radio Engineers. In 1996,
he received the 1995 J. K. Mitra Award of the Institution of Electronics
and Telecommunication Engineers (IETE) for the best research paper.
Since 1981 he has published over 170 technical papers in books, journals and
conferences, mainly in the field of microwave computer-aided design.
He is the president of the IEEE 2000 International Z\"urich
Seminar on Broadband Communications (IZS'2000) and President of the
EMC Congress in Z\"urich. He is an Associate Editor of the IEEE Microwave
and Wireless Components Letters and a member of the Editorial Board
of the IEEE Transaction on Microwave Theory and Techniques.
Since 1992, he serves also on the Technical Program Committee of the IEEE
International Microwave Symposium, the
MTT-S Technical Committee on Microwave
Field Theory and in 1999 on the Technical Program Committee of the
European Microwave Conference. He is the chairman of the IEEE Swiss
Joint Chapter on IEEE MTT-S, IEEE AP-S, and IEEE EMC-S.


\begin{thebibliography}{1}
\bibitem{0}
K. Tomizawa, {\em{Numerical simulation of submicron semiconductor
devices}}. Artech House: London, Boston, 1993.
\bibitem{1}
K. Blotekjaer,"Transport equations for electrons in two-valley
semiconductors,"
{\em{IEEE Trans. Electron Dev.}}, vol. 12, pp. 38-47, 1970.
\bibitem{6}
D.M. Caughey, R.E. Thomas, "Carrier mobilities in silicon empirically
related to doping and field,"
{\em{IEEE Proc.}}, vol. 55, pp. 2192-2193, 1967.
\bibitem{2}
Y.K. Feng, A. Hintz, "Simulation of submicrometer GaAs MESFETs using
a full dynamic transport model," {\em{IEEE Trans. Electron Dev.}},
vol. 35, pp. 1419-1431, 1988.
\bibitem{3}
M.A. Alsunaidi, S.M. Hammadi, S.M. El-Ghazaly,
"A parallel implementation
of a two-dimensional hydrodynamic model for microwave semiconductor device
including inertia effects in momentum relaxation,"
{\em{Int. J. Num. Mod.: Netw. Dev. Fields}}, vol. 10, pp. 107-119, 1997.
\bibitem{4}
G. Baccarani, M.R. Wordemann, "An investigation of steady-state
velocity overshoot in silicon,"
{\em{Solid-State Electron.}}, vol. 28, pp. 407-416, 1985.
\bibitem{8}
A. Gnudi, F. Odeh, M. Rudan, "Investigation of nonlocal
transport phenomena in small semiconductor devices,"
{\em{European Transactions on
Telecommunications and Related Technologies}}, vol. 1, no.3, pp. 307-312,
1990.
\bibitem{7}
C. Canali, C. Jacoboni, G. Ottaviani, A. Alberigi Quaranta, "High-field
diffusion of electrons in silicon,"
{\em{Appl. Phys. Lett.}}, vol. 27, pp. 278-280, 1975.
\bibitem{9}
D.L. Scharfetter, H.K. Gummel, "Large-signal analysis of a silicon Read 
diode oscillator,"
{\em{IEEE Trans. Electron Dev.}}, vol. 16, no.1, pp. 64-77, 1969.
\bibitem{10}
T. Tang, "Extension of the Scharfetter-Gummel algorithm to the energy
balance equation," {\em{IEEE Trans. Electron Dev. 1984}}, vol. 1,
no. 12, pp. 1912-1914, 1984.
\bibitem{5}
C.L. Gardner, "Numerical simulation of a steady-state electron shock wave
in a submicrometer semiconductor device,"
{\em{IEEE Trans. Electron Dev.}}, vol. 38, pp. 392-398, 1991.
\end{thebibliography}
\end{document}